\documentclass[10pt]{article}

\usepackage{amsmath}
\usepackage{amssymb}

\usepackage{graphicx}

\usepackage{cite}

\usepackage{color}

\usepackage[labelfont=bf,labelsep=period,justification=raggedright]{caption}
\usepackage{hyperref}

\newcommand{\avg}[1]{{\langle #1\rangle}}

\newcommand{\cH}{\mathbf{H}}
\newcommand{\cG}{\mathbf{G}}
\newcommand{\cU}{\mathbf{U}}

\makeatletter
\renewcommand{\@biblabel}[1]{\quad#1.}
\makeatother

\date{}

\begin{document}

\begin{center}
{\Large
\textbf{Empirical studies on the network of social groups: the case of Tencent QQ}
}
\\
~
\\
Zhi-Qiang You$^{1}$,
Xiao-Pu Han$^{1,\ast}$,
Linyuan L\"u$^{1}$,
Chi Ho Yeung$^{2}$
\\
$^{1}$Alibaba Research Center for Complexity Sciences, Hangzhou Normal University, Hangzhou 311121, China\\
$^{2}$Department of Science and Environmental Studies, The Hong Kong Institute of Education, Hong Kong\\
$\ast$ E-mail: Corresponding author: xp@hznu.edu.cn

\end{center}

\section*{Abstract}

Participation in social groups are important but the collective behaviors of human as a group are difficult to analyze due to the difficulties to quantify ordinary social relation, group membership, and to collect a comprehensive dataset. Such difficulties can be circumvented by analyzing online social networks.
In this paper, we analyze a comprehensive dataset obtained from Tencent QQ, an instant messenger with the highest market share in China. Specifically, we analyze three derivative networks involving groups and their members -- the hypergraph of groups, the network of groups and the user network -- to reveal social interactions at microscopic and mesoscopic level. Our results uncover interesting behaviors on the growth of user groups, the interactions between groups, and their relationship with member age and gender. These findings lead to insights which are difficult to obtain in ordinary social networks.





\section*{Introduction}

Social interactions are essential to us, yet our understanding on their collective behaviors is limited. Major reasons include the difficulties to quantify individual social relationship and to collect a comprehensive dataset. Nevertheless, the rapid development of the Internet has revolutionize the form of social interactions from postal mails, telephone voice calls, physical meeting and gathering, to emails, instant messaging, online forum and online social networks. Through the internet, interactions are quantified into data which greatly facilitates the studies of social networks. Many exciting findings are revealed. As an example, the hypothesis of six degrees of separation was initialized in 1930s~\cite{karinthy29}, which states that any two person can be connected to by a small number of acquaintances, was only recently tested on Facebook network which gives an average degree of separation of roughly 4~\cite{Backstrom}. Other features revealed on online social networks include power-law degree distribution~\cite{Albert}, community structure~\cite{Girvan, Newman2004} and special communication patterns~\cite{Rybski2009,Rybski2012,JiangZQ}.

So far the studies on online social networks focus mainly on individual social relationship, leaving another important aspect -- participation in social groups -- less understood. It is because the collective behavior of human as a group is difficult to study in ordinary social networks due to the ambiguity in quantitatively affiliating individuals to specific groups. This problem does not occur on the Internet since group-based applications have a definite membership identity for individuals. For instance, prototype online applications such as chatrooms and bulletin board systems (BBS) involve individual users joining and posting messages where membership identity is well defined~\cite{Goh,WangP2014}. These applications set the basis for existing social applications and instant messengers include Windows Live and Google messengers, Whatsapp, Skype, Fetion and Tencent QQ.In these applications, users  create social groups on demand and lead to social networks which are more extensive and complicated than their physical counterparts.

Two different types of online social groups can be formed on the Internet. The first one is similar to ordinary social networks, which are joined by friends with real personal relationship. Circles in Google plus, Skype and Whatsapp groups belong to this type~\cite{Kairam}. The second one is more unique to online social networks, consisting of groups of individuals with common interests but without prior personal relationship, for instance, membership in forums and student bulletins. They connect individuals beyond ordinary social networks and extend the social scope of individuals. Despite the difference in their nature, the two types of networks are interdependent on each other~\cite{Havlin2010,Havlin2012,Havlin2013}. For instance, two users in the same forum may become close personal friends and participate in each others' personal social networks. By making new friends, an individual may find new interests and join new forums. This new form of social evolvement is unique to online social networks and has greatly supplemented or even replaced its physical counterpart.

In this paper, we analyze a comprehensive dataset obtained from Tencent QQ, an instant messenger with the highest market share in China. Both types of social networks are established on QQ and the interactions between the two networks are expected. Specifically, we analyze three different networks involving groups and their members -- the hypergraph of groups, the network of groups, and the user network -- to reveal social interactions at microscopic and mesoscopic levels. Our results uncover interesting behaviors on the growth of user groups, the interactions between groups, and their relationship with member age and gender. These findings reveal unique phenomenon in online social networks, as well as insights which are otherwise inaccessible in ordinary social networks.

\section*{Method}
\subsection*{Data description}
Tencent QQ (commonly abbreviated as QQ\footnote{The website of Tencent QQ: \url{http://www.qq.com}}) is an instant communication tool developed by Tencent Holdings Limited in 1999. To date, it has over 700 million active users and has become the largest online application in China. QQ users can send messages, share photos and files, post microblogs, and voice or video chat with friends using computers or smartphones.

Social group is one of the main features of QQ which allows multiple users to communicate instantly. A message posted by a member is immediately received by all the other group members. When necessary, any two members can communicate via individual channel. Depending on the activeness of a user, each of them can create no more than six groups.
Groups can be searched by their ID's or names and other users can join the group upon the approval by the administrator, i.e. the group creator. QQ limits the group size by 100, 200, 500, and 1000, depending on the activeness of the group creator. Other than personal relationships, some groups are formed by members with common interests, e.g. movies, or belong to the same organizations, e.g. universities or companies.
The latters are usually exclusive social circles based on physical organizations.

The QQ dataset \cite{QQdata} we examine covers more than 58,523,451 groups and 274,335,675 users, of which 48,665,873 groups has the information with all ID, member list, and date. Due to the limit of 2000 groups which are allowed for an ordinary user to join, 34 users who joined more than 2000 groups are considered as robots and are excluded from our analyzes. Since some users do not indicate his/her gender or age, or provide some seemingly false information, e.g. 0 year old, we exclude users without gender information or younger than 10 or older than 70. Overall, there are 273,205,008 users with gender information, of which 42.5$\%$ (116,136,202) are females, and 244,521,767 users with age between 10 and 70. For most of the QQ groups,
its ID, its member list with gender and age, and the date of which it was established are known. The oldest and the youngest groups in our dataset are formed on 22$^{\rm nd}$ September, 2005 and 25$^{\rm th}$ March, 2011, respectively. We thus only use data up to 25$^{\rm th}$ March, 2011.

\subsection*{Networks construction}
\begin{figure}[!ht]
\begin{center}
 \centerline{\includegraphics[width=6in]{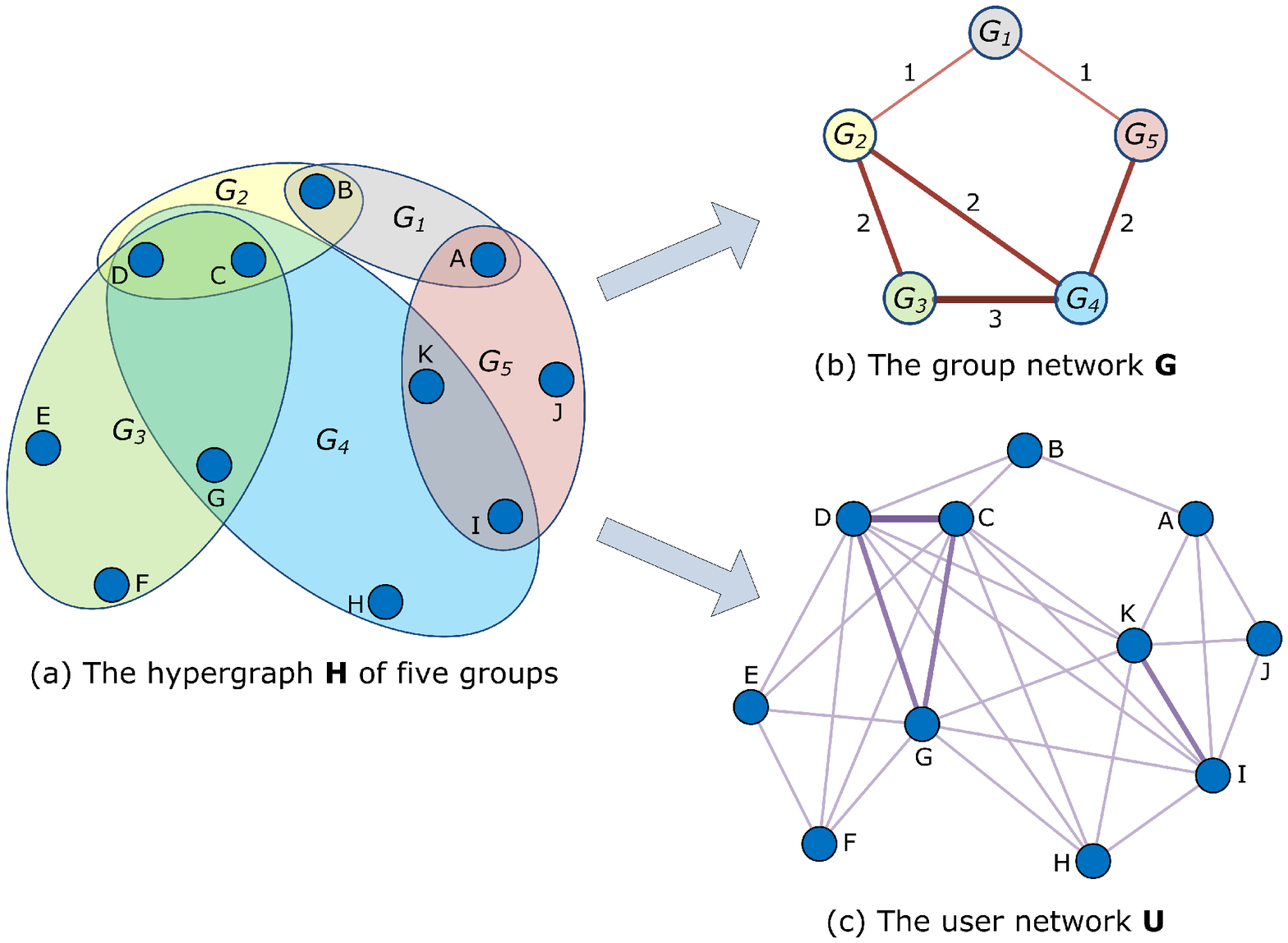}}
 \end{center}
  \caption{{Schematic diagram showing (a) the user-group hypergraph $\cH$, (b) the group network $\cG$, and (c) the user network $\cU$.} The data is composed of five groups denoted by the colored ellipses in (a) and eleven users. The thickness of edges in (b) and (c) is proportional to the weight on the edges.
  }
  \label{hypergraph}
\end{figure}

We examine three types of networks embedded in the datasets:
\begin{enumerate}
\item
\emph{User-group hypergraph} -- A hypergraph~\cite{Berge1,Berge2} is a graph of nodes and hyperedges each of which connects two or more nodes. As shown in Fig.~\ref{hypergraph}(a), the hypergraph in our dataset describes the user-group relationship with nodes representing individual users and hyperedges representing groups. For instance, user $B$ is a member of group $G1$ and $G2$, and are thus connected to $A$ via hyperedge $G1$ as well as to $C$ and $D$ via hyperedge $G2$. In this paper, we label the results obtained on the user-group hypergraph by superscript $\mathbf{H}$.
\item
\emph{Group network} -- As shown in Fig.~\ref{hypergraph}(b), group networks in our context refer to weighed networks where nodes represent individual groups, and two groups are connected if they have at least one common member. The weight on the edge is defined as the number of common users between the two groups. For instance, group $G3$ and $G4$ in Fig.~\ref{hypergraph}(a) have 3 common users, the edge connecting $G3$ and $G4$ in Fig.~\ref{hypergraph}(b) has a weight of 3. In this paper, we label the results obtained on the group network by superscript  $\mathbf{G}$.
\item
\emph{User network} -- To focus on the behaviors of social groups, the user network in our context is not the ordinary friendship network in QQ, but instead is a weighted network which only connects two users if they are members of at least one common group. Hence, all members in a group are fully connected to each other. The weight of an edge connecting a pair of users is equal to the number of groups they both join. As shown in Fig.~\ref{hypergraph}(a), both user $C$ and $D$ are members of $G2$, $G3$ and $G4$, and hence the weight on the edge connecting user $C$ and $D$ in Fig.~\ref{hypergraph}(c) is 3. In this paper, we label the results obtained on the user network by superscript  $\mathbf{U}$.
\end{enumerate}

The notations used throughout the paper are summarized in Table~\ref{notations}. 

\begin{table}[htbp]
\caption{Notations in the paper.}\label{notations}
\begin{tabular}{llp{13cm}}
\hline
Notations && Description\\
\hline
$s^{\cH}$ &&The size of group $H$, namely the total number of users in a group\\ 
$\avg{s^{\cH}}$ &&The average value of size of groups \\
$k^{\cH}$ && The number of groups that a user joined, the node's hyperdegree\\ 
$\avg{k^{\cH}}$ && The average value of node's hyperdegree\\
$k_{max}^{\cH}$  &&The maximum value of joined groups of users in a group\\ 
$k^{\cG}$&& The node degree in group network, namely the number of connected groups of a group \\ 
$K^{\cG}$&& The weighted node degree in group network, namely the number common users between a group and others\\ 
$w^{\cG}$&&The edge's weight in group network, which is defined as the number of common users between two groups\\ 
$d^{\cG}$&&The distance between groups in group network\\ 
$\avg{d^{\cG}}$ &&The average distance between groups in group network\\ 
$C^{\cG}$ &&The local clustering coefficient in group network\\
$\avg{C^{\cG}}$ &&The average value of local clustering coefficient in group network\\
$k^{\cU}$ &&The user's degree in user network, for an arbitrary user $i$, $k^{\cU}(i)$ is the number of users who are in the same groups with $i$\\ 
$\avg{k^{\cU}}$ &&The average user's degree in user network\\
$w^{\cU}$ &&The edge's weight in user network, which is defined as the number of common groups between two users\\ 
$d^{\cU}$ &&The distance between users in user network\\ 
$\avg{d^{\cU}}$ &&The average distance between users in user network\\ 
$a_i$&& The age of user $i$\\ 
$\avg{a}$&& The average value of user's age\\
$P(a)$&& The distribution of users' ages\\ 
$c_{va}$ &&The coefficient of variation of age, $c_{va}=\sigma_a/\avg{a}$ where $\sigma_a$ is the standard deviation of member age\\
$c_{vu}$ &&The coefficient of variation of the number of neighboring users, $c_{vu}=\sigma_u/\avg{k^{\cU}}$ where $\sigma_u$ is the standard deviation of member degree in user network\\
$\avg{c_{va}}_f$ &&The average coefficient of variation of neighboring users' ages\\
$\avg{c_{vu}}_f$ &&The average coefficient of variation of neighboring users' degree\\
\hline
\end{tabular}
\end{table}

\section*{Results}

\subsection*{The Structural Properties of the User-Group Hypergraph $\mathbf{H}$}

\begin{figure}[!ht]
\begin{center}
 \centerline{\includegraphics[width=6in]{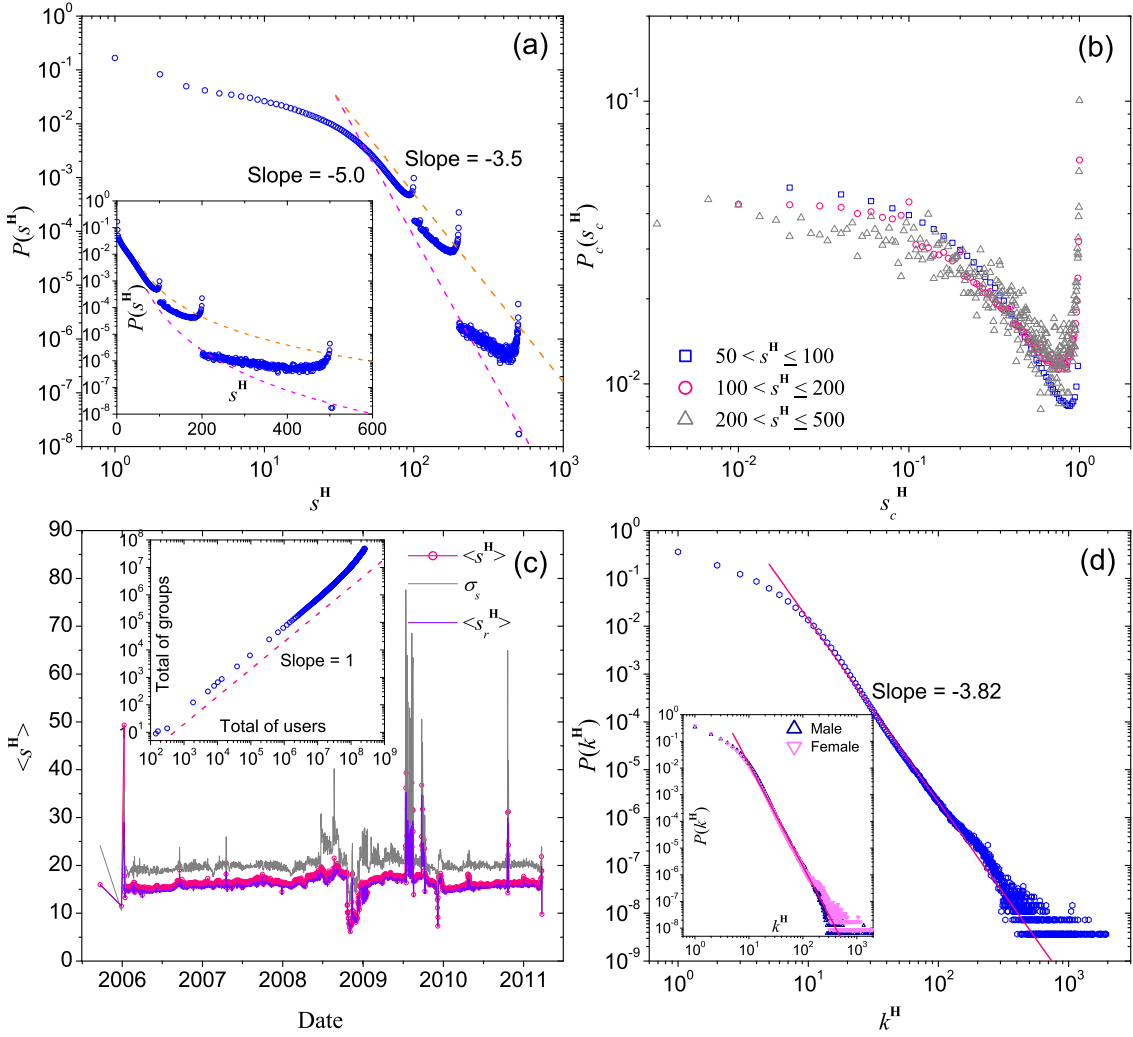}}
 \end{center}
  \caption{
  {Statistics for the hypergraph $\cH$.}
  (a) $P(s^{\cH})$, the distribution of group size $s^{\cH}$, with the distribution in semi-log scale shown in the inset. The two dashed lines in show the range of the tail exponent of $P(s^{\cH})$, namely  $-3.5$ (orange) and $-5.0$ (magenta). (b) The data collapse of the different broken parts on $P(s^{\cH})$ after re-scaling, in which $s^{\cH}_c=s^{\cH}/\avg{s^{\cH}}$, here $\avg{s^{\cH}}$ is the average value of $s^{\cH}$ in each section, and $P_c(s^{\cH})$ is the corresponding re-scaled probability.
  (c) The average (pink) and the standard deviation (grey) of group size given specific date of establishment, and the inset shows the scaling relationship between total of groups and total of users at each date. (d) The distribution $P(k^{\cH})$ of the number of joined group by individual users. $P(k^{\cH})$ for male and female users are shown in the inset. The pink lines correspond to power-law fits with exponent $-3.82$.}\label{hyperdegree}
\end{figure}

The distribution of social group size is one of the most interesting features in a social network. We remark that group size is equivalent to the rank $s^{\cH}$ of the corresponding hyperedge in the user-group hypergraph $\mathbf{H}$. As we can see in Fig.~\ref{hyperdegree}(a), the distribution $P(s^{\cH})$ shows a slow and smooth decay in the range $0\le s^{\cH}\le 50$. The decay becomes faster for $s^{\cH}> 50$ and the curve becomes discontinuous at $s^{\cH}=100, 200, 500$ and $1000$, due to the limitation of group size by QQ. We find that the broken parts of the curve can be enclosed by two power laws with exponent $-3.5$ and $-5.0$, i.e. the two dashed lines in Fig. \ref{hyperdegree}(a). These exponents are more negative than similar exponents observed in other social networks, suggesting that it is more difficult for a group to maintain a large member community than for an individual to maintain a large number of friends. The results indicate a more homogeneous nature in the distribution of group size,probably because maintaining such close relationship in a large group, e.g. clubs or organizations, is not easy, which limits the growth of group. On the other hand, we show in Fig. \ref{hyperdegree}(b) a data collapse of the different broken parts after re-scaling, implying that formation mechanisms of groups are similar regardless of their size.

Other than the distribution of group size, Heaps' law is another universal law that govern disparate systems \cite{Heaps}. It has been pointed out that the Heaps' law can be considered as a derivative phenomenon if the system obeys the Zipf's law. The relation between the exponents of Zipf's law and
Heaps' law was given by L\"u et al. \cite{LYPLOS}. The inset of Fig. \ref{hyperdegree}(c) shows the relation between the number of groups and the number of users, which follows heaps' law with exponent equals to 1.

Intuitively, we expect older groups to have a larger size since they have a longer time to accumulate members. To reveal the correlation between the size of a group and the date of which it is formed, we compute the average size of groups established on the same date. As we can see in Fig. \ref{hyperdegree}(c), the average size $\avg{s^{\cH}}$ is almost independent of the date of establishment, which is contrary to our belief. This result may imply that most groups do not grow significantly after establishment, and the group size is mainly determined by the number of users who joined the group shortly after the group was created. It is because when a group is created, its information usually spreads rapidly in the creator's social circle. As a result, most interested users join the group once they heard about it.
Occasionally, a small number of users may join existing groups but on the other hand, some existing members may leave the group leading to an equilibrium group size. This certainty on group members creates some difficulties into the studies on recommendation algorithms for QQ groups.

The above pictures are further supported by the standard deviation $\sigma_s$ of group size, which again does not increase with the age of a group. Moreover, excluding the groups that is close to the size limits ($s^{\cH}$ in the range 90 - 100, 180 - 200, 450 - 500), the average size $\avg{s^{\cH}_r}$ of the remaining groups also shows the same phenomenon (the violet curve in Fig. \ref{hyperdegree}(c)). This observation of constant size is different from many other slow-growing social networks.

Other than the group size, the number of groups joined by an individual user is also an important characteristic of a social network. In the context of hypergraph, one can represent the number of group joined by a user by the hyperdegree $k^{\cH}$ of the user. Figure \ref{hyperdegree}(d) shows the distribution $P(k^{\cH})$ with a tail well fitted by a power law with exponent $-3.82$. Although the exponent is more negative  than most of the other social networks, a power-law decay does imply that users which joined a large number of groups are present. Unlike previous studies which revealed differences based on gender, we observed similar $P(k^{\cH})$ for both male and female users.

\begin{figure}[!ht]
\begin{center}
 \centerline{\includegraphics[width=6in]{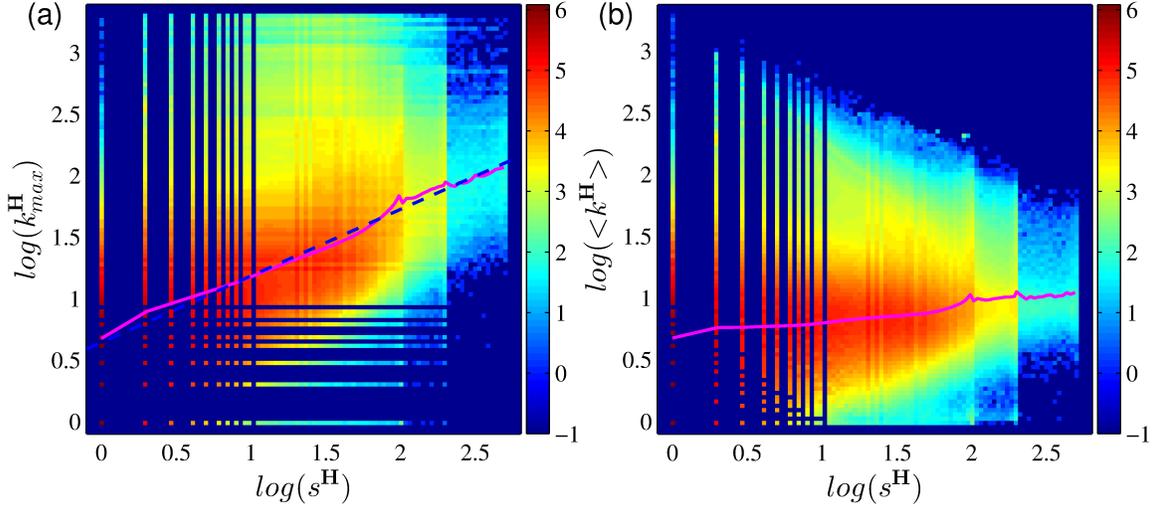}}
 \end{center}
  \caption{{The heat maps showing the correlation between group size and the number of joined group of members.} The color scale corresponds to the log-frequency of occurrence between the size of a group and (a) the largest number of group joined by an individual member in the group, and (b) the average number of group joined by the members in the group. The pink lines show the curves on their means along vertical values, and the blue dashed line in (a) shows the fitting power function with slope $0.54$.
  }
  \label{kH}
\end{figure}

Denote by $k^{\cH}_{max}$ the largest number of joined groups by an individual member in a group, we observe a strong positive correlation between $k^{\cH}_{max}$ and the group size $s^{\cH}$ as shown in Fig. \ref{kH} (a). The dependence of $\avg{k^{\cH}_{max}}$ on $s^{\cH}$ can be fitted by a power law with exponent 0.54. The increasing $\avg{k^{\cH}_{max}}$ with group size may not seem surprising since active users are proportionately more likely to be present in larger groups which include more users. This is true even if active users do not have a preference for joining large groups. This is further supported by Fig. \ref{kH}(b), which shows that the average $k^{\cH}$ among group members does not increase with group size, suggesting a similar composition of active and less active member users across groups of different sizes.

\subsection*{The Structural Properties of Group Network $\mathbf{G}$}

After examining the marcoscopic characteristics of groups, we move on to reveal their microscopic interactions. In this respect, the weighted group network characterizes an indirect interaction between groups when they share some common members. As a reminder, two groups are considered connected if they share some common neighbors and the weight of the edge is the number of users who joined both groups.

\begin{figure}[!ht]
\begin{center}
 \centerline{\includegraphics[width=6in]{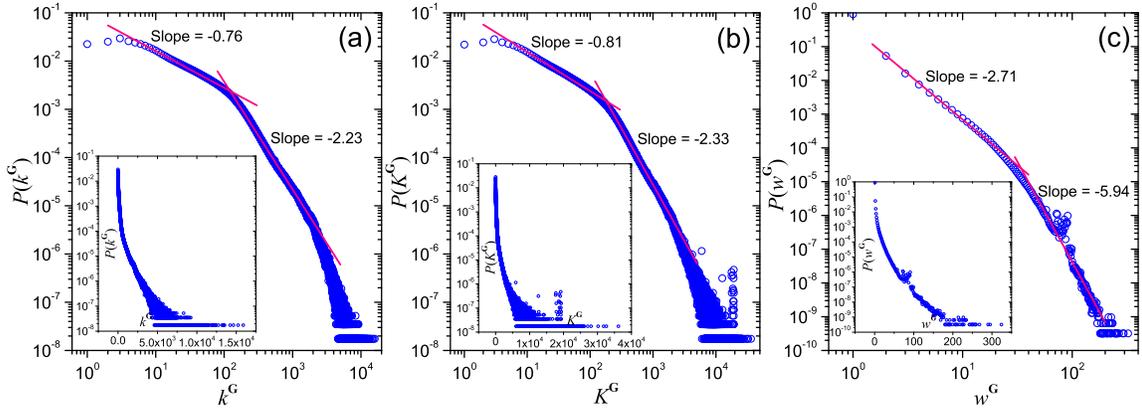}}
 \end{center}
  \caption{{Properties of the group network $\cG$.} The figures show (a) the distribution $P(k^{\cG})$ of group degree, (b) the distribution $P(K_{\cG})$ of weighted group degree $K_{\cG}$ of G, and (c) the distribution $P(w^{\cG})$ of edge weight. The insets show the same curves in semi-log scale. }
  \label{group}
\end{figure}

As shown in Fig.~\ref{group}(a), the distribution $P(k^{\cG})$ of the group degree $k^{\cG}$ shows a power law with exponent $-0.8$ when $k^{\cG}<120$ and another power law with exponent $-2.23$ when $k^{\cG}>120$. Similarly, as shown in Fig.~\ref{group}(b), the weighted degree distribution $P(K^{\cG})$ shows a power law with exponent $-0.81$ when $K^{\cG}<160$ and another power law with exponent $-2.33$ when $K^{\cG}>160$. The results imply that a group only share members with a small number of groups, usually at most of the order $O(10^2)$ among the 58 million groups in the QQ network.
On the other hand, the number of common members between a pair of groups, i.e. the weight of edge, also obeys a two-region power-law as shown in Fig.~\ref{group}(c), with an exponent $-5.94$ at the tail. This implies that the number of users who have interests in a common pair of groups are limited to the order of $O(10^2)$.

\begin{figure}[!ht]
\begin{center}
 \centerline{\includegraphics[width=6in]{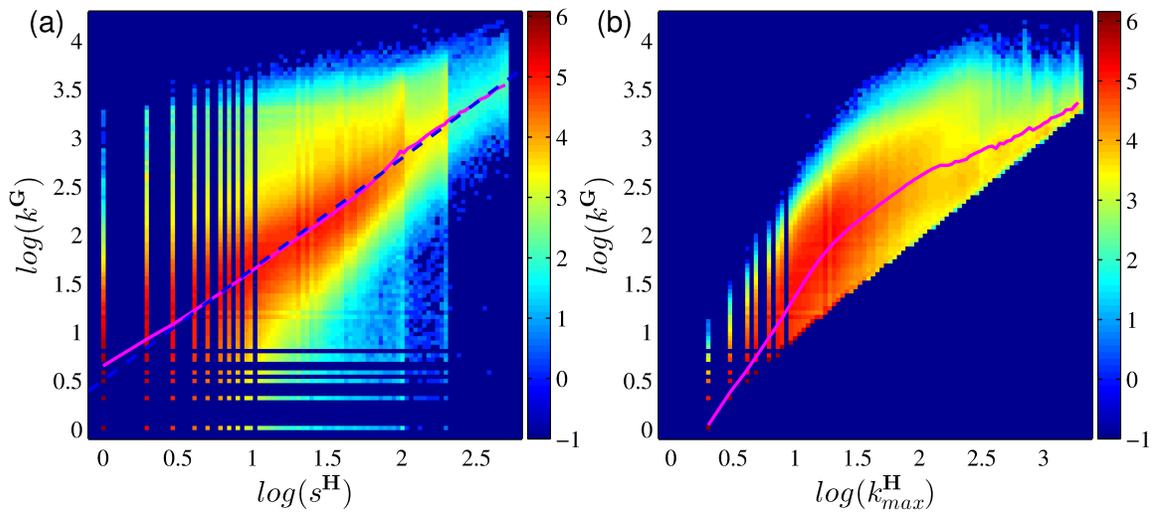}}
 \end{center}
  \caption{{The heat maps which show the correlation (a) between $k^{\cG}$ and $s^{\cH}$ (a), and (b) between $k^{\cG}$ and $k^{\cH}_{max}$.} The color scale corresponds to the log-frequency of occurrence. The pink lines show the curves on their means along vertical values, and the blue dashed line in (a) shows the fitting power function with slope $1.14$.
  }
  \label{pattern}
\end{figure}

The degree of a group is dependent on two factors, namely (i) the number of users in the group, and (ii) the total number of other groups joined by its members. Figure~\ref{pattern}(a) shows the relation between the group degree $k^{\cG}$ and the group size $s^{\cH}$, such that the relation between $s^{\cH}$ and the corresponding $\avg{k^{\cG}}$ is given by the pink curve.
The results show that group degree increases with group size, which is expected since the number of different groups joined by the members of a larger group should be proportionately higher. In Fig.~\ref{pattern}(b), a similar statistics shows the relation between the group degree $k^{\cG}$ and $k^{\cH}_{max}$, the largest number of joined groups by an individual member in a group. The reason is similar to that in Fig.~\ref{pattern}(a), since a larger group has proportionately more active members, the largest number of group joined by an individual member is higher. The average of $k^{\cG}$ has an obvious transition from a faster growth to a  slower growth (see Fig. \ref{pattern}(b)), indicating that $k_G$ is more strongly dependent  on $k^{\cH}_{max}$ when $k^{\cG}$ is smaller than 100. This results imply that when the group degree $k^{\cG}$ is small, the active users have a significant role in improving $k^{\cG}$.

\begin{figure}[!ht]
\begin{center}
 \centerline{\includegraphics[width=5in]{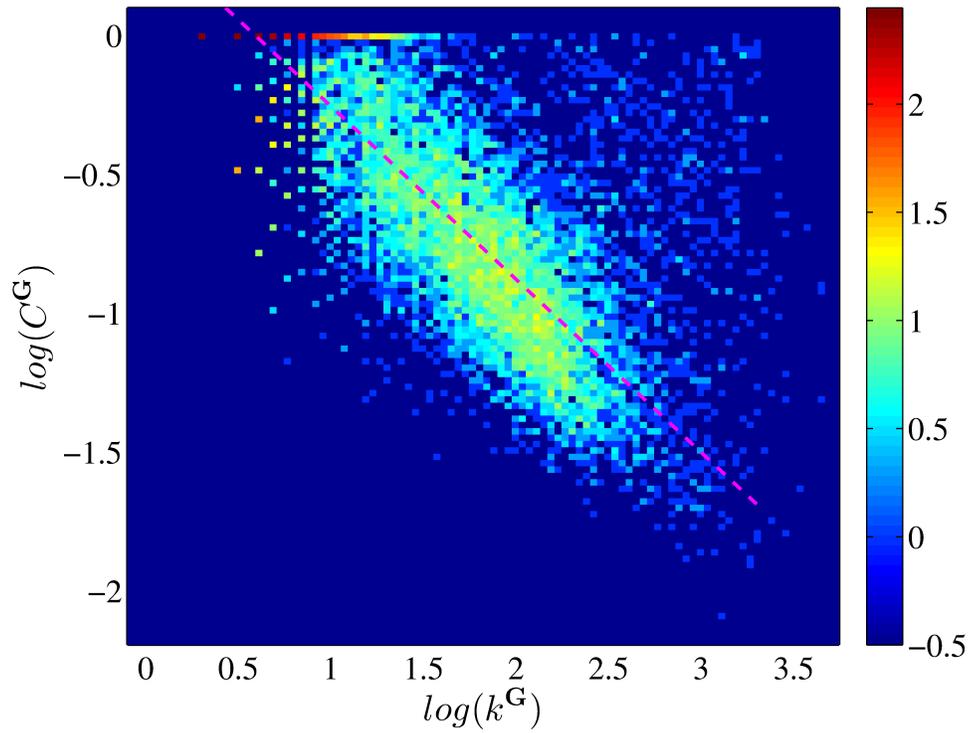}}
 \end{center}
  \caption{{The heat map which shows the correlation between local clustering coefficient $C^{\cG}$ and the degree $k^{\cG}$ in group network $\cG$.} The color scale corresponds to the log-frequency of occurrence over $10^4$ randomly sampled groups.
  The pink dashed line shows the fitting curve with slope $-0.62$ on the means along vertical values.
  }
  \label{cluster}
\end{figure}

Finally, we show that the QQ group network is sparse but shows ``small-world" phenomenon, similar to the friendship network of Facebook~\cite{Backstrom, Ugander}. Comparing to Facebook, the average degree and average weighted degree are slightly smaller in QQ group network, with values $108.8$ and $133.6$ respectively. These degrees are small given the large size of the network, indicating the network is sparse. To show the ``small-world" phenomenon, we randomly sample $2 \times 10^4$ pairs of groups and remarkably find that their average distance is only $3.7$, similar to the four degrees of separation observed in Facebook~\cite{Backstrom}. We also compute the local cluster coefficient $C^{\cG}$
\begin{align}
	C^{\cG}=\frac{2n_T}{k^{\cG}(k^{\cG}-1)}
\end{align}
for $10^4$ random chosen groups, such that $n_T$ is the number of connection among the neighbors of the group. We show the frequency of the values $(k^{\cG}, C^{\cG})$ for individual group in Fig.~\ref{cluster}. As we can see, $C^{\cG}$ is negatively related to $k^{\cG}$ in a rough power-law relation with exponent $-0.62$, which is similar to the Facebook case~\cite{Backstrom}. The average value of $C^{\cG}$ is 0.35, which is high compared to the other social networks.

\subsection*{The Structural Properties of User Network $\mathbf{U}$}

\begin{figure}[!ht]
\begin{center}
 \centerline{\includegraphics[width=6in]{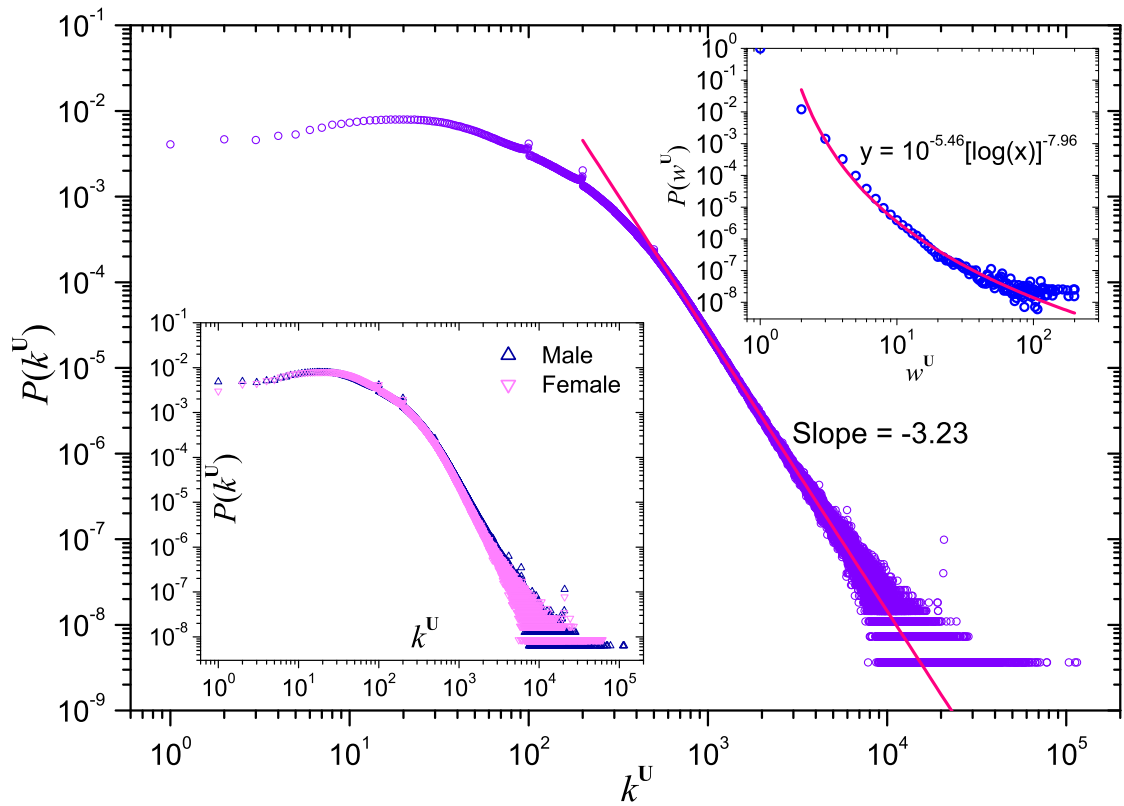}}
 \end{center}
  \caption{{Degree distribution $P(k^{\cU})$ in the user network $\cU$.} The bottom inset shows the same distribution over male and female users respectively. The top inset shows the distribution $P(w^{\cU})$ of edge weight $w^{\cU}$.}
  \label{user}
\end{figure}

A similar analysis is conducted for the weighted user network. We show in Fig.~\ref{user} the degree distribution $P(k^{\cU})$, which has a power-law tail with exponent $-3.22$, and an average value $135.3$. The degree distributions for male and female users do not show obvious difference and are shown in the bottom inset of Fig.~\ref{user}. By averaging $10^4$ random pairs of users, we compute the average distance between a pair of user to be $4.17$, similar to the distance observed in Facebook friendship network. These results show that the user network is sparse and exhibits a small-world phenomenon. By comparing the degree distribution $P(k^{\cU})$ and the weighted degree distribution $P(w^{\cU})$ as shown in the top inset of Fig.~\ref{user}, we observe that the latter can be fitted well by a decay function $P(w^{\cU})\approx 10^{-5.45}[\log(w^{\cU})]^{-7.96}$, which is slower than power-law.
It implies that users with large degree are more likely to share groups with other users, resulting in a large edge weight, and thus a shift of the tail part to the right.

\subsection*{Grouping behaviors and user age}

\subsubsection*{The most active age group in QQ group participation}

To make the best use of the available data, we go on to reveal the relation between user age and their joined groups. Similar studies have shown that the preference for gender in social contacts changes along with age \cite{Palchykov}. Here we will reveal similar changes in grouping preference along with age.

Figure \ref{age}(a) shows the distribution of individual user age, namely $P(a)$, and the distribution of average member age of groups, namely $P_G(\avg{a})$. As we can see, members in QQ-groups are mainly young users of  around 20 years old. As shown in Fig.\ref{age} (b), the distributions $P(k^{\cH})$ of the number of group joined by an individual user, i.e. the hyperdegree $k^{\cH}$ in the user-group hypergraph, is slightly dependent on ages: comparing different ages, the decay of $P(k^{\cH})$ for users in the range of 40 to 44 is faster in the small $k^{\cH}$ regime and slightly slower in the large $k^{\cH}$ regime. We further show that the number of joined group $\avg{k^{\cH}}$ is highest at two distinct ages, showing a bimodal form, where the first peak is located at $a \approx 15$ and corresponds to a group of teenagers, and the second peak should appear in $a>65$ and corresponds to the elderly. The number of joined group is minimum when users are at their 40s.

These results indicate that both teenagers and the elderly are active in group-based social interactions, in contrast to the less active middle-aged users at their 40s. We examined  the groups joined by several elderly users and find that majority of them are groups for entertainment, indicating  their needs for leisure activities and social interactions. For the users at their 40s, they are likely to be engaged either in family or works and are thus less active in joining QQ groups.

\subsubsection*{The distribution of member age in groups}

\begin{figure}[!ht]
\begin{center}
 \centerline{\includegraphics[width=6in]{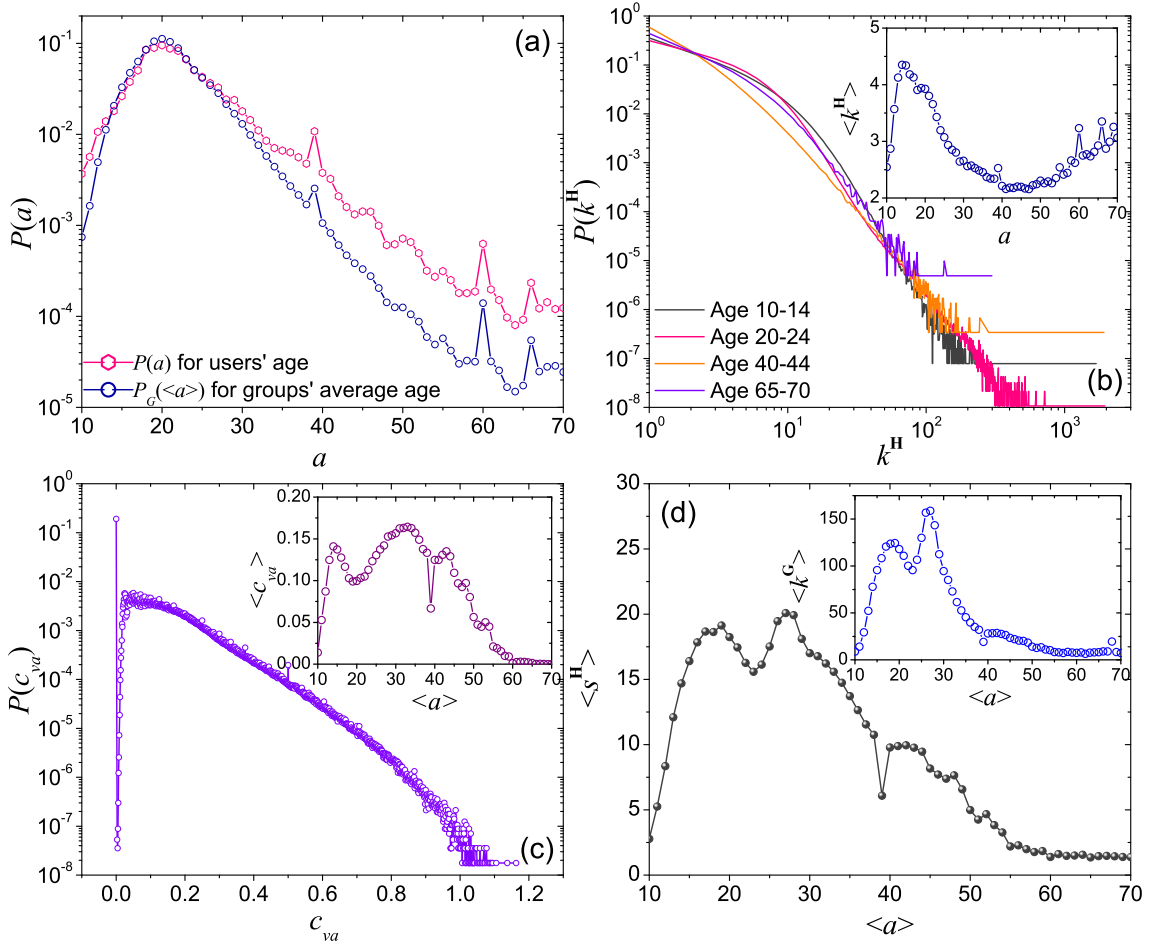}}
 \end{center}
  \caption{{The relation between member age and group characteristics.} (a) The distribution of age over individual users and average member age over individual groups. (b) The distribution of the number of joined groups by different age groups. Inset: the average value of of joined group over users at different ages. (c) The distribution $P(c_{va})$ of the coefficient of variation $c_{va}$ for users' age in each group, and the average value $\avg{c_{va}}$ as the function of the average age of groups is shown in the inset.
   (d) The dependence of average group size $\avg{s^{\cH}}$ on average group member age $\avg{a}$. Inset: The dependence of average group degree $\avg{k^{\cG}}$ on average group member age $\avg{a}$.
   }\label{age}
\end{figure}

We compute the coefficient of variation $c_{va} = \sigma_a/ \avg{a}$ of each group, where $\sigma_a$ and $\avg{a}$ are the standard deviation and the average of the member age in the group respectively. The distribution of $c_{va}$ is shown in Fig. \ref{age}(c), which shows that the values of $c_{va}$ in more than half of the groups are smaller than 0.1, indicating group members are usually of similar age. This result is expected since users with similar age usually have similar interests or are engaged in similar institutes and thus are more likely to meet each other.

On the other hand, the behaviors of $c_{va}$ for groups with different average ages are different. As shown in Fig. \ref{age}(c), $c_{va}$ is first peaked for groups with average age $\avg{a} \approx 14$, corresponding to groups of teenagers, and also peaked for groups with age $\avg{a} \approx 33$, corresponding to groups of adults who are likely to be at the intermediate level of their careers. The average value $\avg{c_{va}}$ for groups with elderly users is low. In general, $\avg{c_{va}}$ can be considered as a measure of the user diversity within the group, and the above findings may imply that teenagers and users at their 30s are more open to make friends with others who may not be in the same age group. On the other hand, the smaller $c_{va}$ for groups with $\avg{a}\approx 20$ may imply users at their 20s are looking for friends who are of similar age, e.g. lovers or fellow university students.

The above interesting bimodal characteristics are observed in the average group size $\avg{s^{\cH}}$ and the average degree $\avg{k^{\cG}}$ of groups in the group network. As shown in Fig.~\ref{age}(d) and its inset, the first peaks of $s^{\cH}$ and $k^{\cG}$ appear at age around 19, while the second one appears at the age of 28. These results reveal a non-monotonic change of group preference with age.

\subsubsection*{Change of group preference with age}

\begin{figure}[!ht]
\begin{center}
 \centerline{\includegraphics[width=6in]{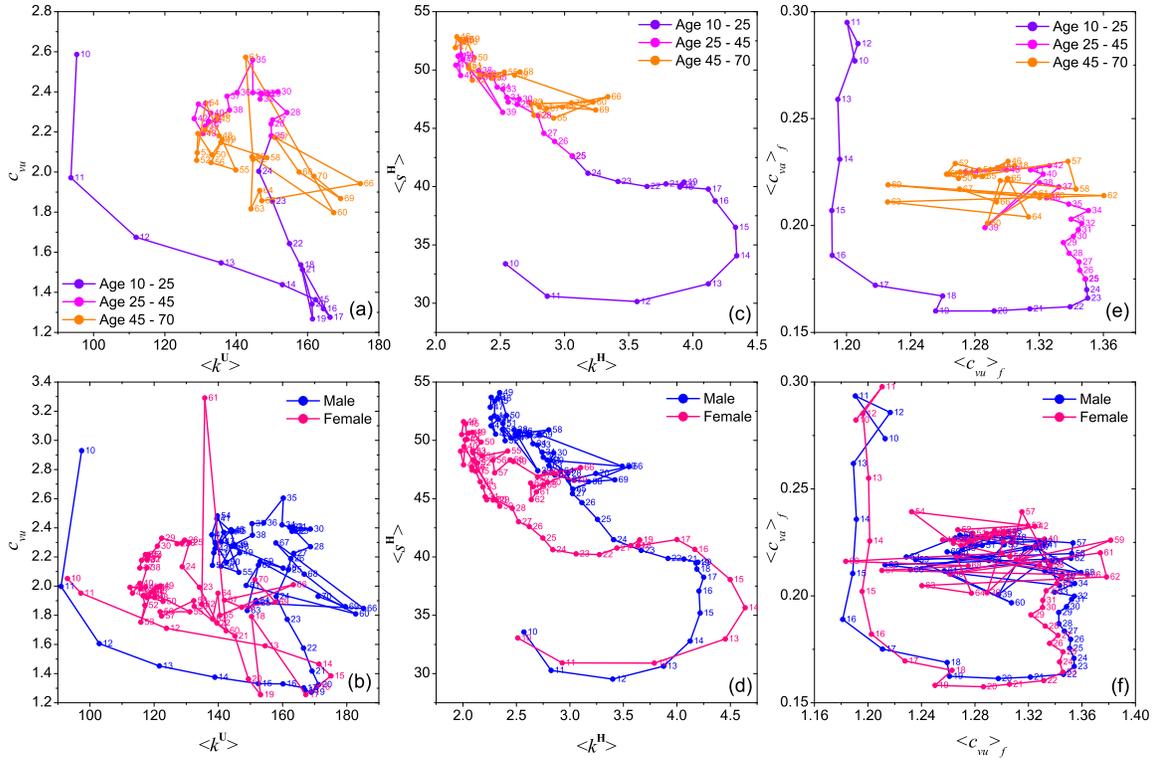}}
 \end{center}
  \caption{
  {The paths of changes of two averaged variables along with age.} (a) X-axis: the averaged value $\avg{k^{\cU}}$ of the degree $k^{\cU}$ of user network $\mathbf{U}$ for users in each age, Y-axis: the coefficient of variation $c_{vu}$ of $k^{\cU}$ for each age. (b) gender differences on the age trail in panel (a).
  (c) Horizontal axis: the average number of joined groups by individual users; Vertical axis: the average value size of the joined groups. (d) The same path in (c) by averaging only male and female users respectively. (e) Horizontal axis: the average coefficient of variation $\avg{c_{vu}}_f$ among the degree of neighbors of a user in the user network $\mathbf{U}$,
   Vertical axis: the average coefficient of variation $\avg{c_{va}}_f$ among the age of neighbors of a user in the user network $\mathbf{U}$.
    (f) The same path in (e) by averaging only male and female users respectively.
     The labels close to each data point corresponds to the value of age, and the different colors in (a), (c) and (d) respectively show the data points in three different age stages. }\label{path}
\end{figure}

To get a clearer picture of the change of group characteristics with age, we (i) compute the average value of variables over groups with particular average member age, and (ii) show simultaneously the change of a pair of variables on a 2D space, which constitutes a path of the group characteristics when average member age increases. As shown in Fig. \ref{path}(a), we show the average user degree $\avg{k^{\cU}}$ in the user network and the coefficient of variation $c_{vu}$, along a path when member age increases.
The results imply that teenagers usually have a smaller but more diverse friendship community until 20 years old, where their friendship community increases in size but decrease in diversity, probably because they are studying in universities. Afterwards, when users start their career, the diversity of friend increases but the friendship community slightly shrinks. These observations are consistent with our previous analyzes which show a transition from a pre-mature regime to a mature regime.

Other than an average over all users, we show a similar path in Fig. \ref{path}(b) by averaging over male or female members in a group. As we can see, the path of the female users shows a faster increase in $k^{\cU}$, i.e. a faster increase in the size of their friendship network, and an earlier transition into the mature regime at the age $a\approx 16$ compared to the age $a\approx 20$ of male users. In general, female becomes mature at an earlier age may  be the reason. In addition, female users show a smaller $k^{\cU}$ in the mature regime compared to male users, reflecting their more significant role in family and thus a limitation on their social contacts.

We continue to show the change of group size $s_{\cH}$ and the number of joined groups $k^{\cH}$ by users at various ages. As we can see in Fig.~\ref{path}(c), a pre-mature regime and a mature regime can be roughly identified in the path, separated at the age around 15. In the pre-mature regime, users tend to join more groups, each with a smaller size, while in the mature regime, users usually join a smaller number of group, each with a larger size. Figure~\ref{path}(d) shows the corresponding path averaged over male and female users only. As we can see, female users show an earlier transition into the mature regime, similar to that observed in Fig. \ref{path}(b). Female users are also observed to join less groups after the transition, which is again consistent with their role in their family.

Finally, we examine  the change in the diversity of the neighboring users' ages and the degree of the neighbors of a user, denoted by $\avg{c_{va}}_f$ and $\avg{c_{vu}}_f$ respectively, with his/her age. As shown in Fig. \ref{path}(e), users of age within the range 15 - 23, i.e. users lying in the transition from the pre-mature to mature regime, usually have a friendship network composed mainly of users of similar age, probably corresponding to a studying stage in colleges. However, the diversity of the neighbors' degree becomes higher  with the increases of age within  the age group from 15 - 23. After the transition, users of age within the range 25-35 usually have a friendship community with wider range of age but similar degree. For elders, they have moderate diversity in terms of both the neighbors' age and degree. Similar path by averaging over male (female) members only is shown in Fig. ~\ref{path}(f).

The above results show that teenagers are generally active in different social communities until 20 years old, where they start their college study, reduce their activities and make friends with fellow universities students. We observe that after the age of 25, the path characterizing different pair of variables enters a mature regime and become stable in a small region of the 2D plane. This may correspond to a transition from the stage of studying to working. In this stage, users tend to join group with greater member diversity. We find that female users are in general less active than male users after the transition into the mature stage, probably because of their important role in their family.

\section*{Discussions}

Participation in social groups is essential, yet our understanding on them is limited due to the difficulties in data collection in ordinary social networks. Fortunately, online social networks do not have such problems. By using a comprehensive dataset obtained from Tencent QQ, we analyzed three derivative networks involving groups and their members. We showed that the distribution of the number of groups joined by an individual follows a power-law, similar to other social networks except a larger decay exponent observed in the present case. The group size of QQ is limited by some specific values, nevertheless, we showed a data collapse on the statistics of groups limited by different maximum size, implying a similar group formation mechanism regardless of their size. Other than distributions, network at the group level shows a small-world phenomenon with an average distance of 3.7. Such findings are remarkable since the there are 58 millions groups in the extremely sparse group network, and yet on average only 3 to 4 steps are required to connect any group pair. All these findings on online social groups are otherwise inaccessible in the studies of their physical counterpart.

To make the best use of available data, we went one step forward to study the interdependence between a group and the age of its members. The results showed a change in the user preference for groups at different ages. A pre-mature and a mature stage can be identified. For youngsters who are still in schools, they are more active in social group participation in QQ and have a larger diversity of friends in terms of age. The situation changes when users are in the age group of 20s, they reduce their activities and make friends with mostly fellow college students. Afterwards, when users start working, they enter the mature stage such that the diversity of their friends and groups increase again. These changes along the growth of age are revealed in various characteristics of their grouping preference.

As we can see, data collected in online social networks has revealed the interaction and participation of users in social groups. The results lead us to a better understanding of social interaction via information technology. Nevertheless, ordinary social interaction is still essential and a comprehensive understanding of the connection between online and ordinary social networks is missing. In this respect, the present study provides useful insights into the study of ordinary social networks, for instance, a guide to the design of surveys and collection of data in  ordinary social networks. We believe our the insights obtained from the present studies are not limited only to online social networks,  but would be useful to fill the missing connection to its physical counterpart.



\section*{Acknowledgments}

This work is supported by the National Natural Science Foundation of China (11205040, 11205042, 11105024, 11305043), and the research startup fund of Hangzhou Normal University, and the EU FP7 Grant 611272 (project GROWTHCOM), and CCF-Tencent Open Research Fund. ZQY acknowledges the Xinmiao Talent Program of Zhejiang Province(Grant No. ZX13005002062). CHY acknowledges the Internal Research Grant of HKIEd.





\newpage


\section*{Figure Legends}

\end{document}